\documentclass[a4paper]{article}
\usepackage{amsmath}
\begin{document}

\title{{\bf On integrability of the Yao--Zeng two-component short-pulse equation }}

\author{
{\sc J.C. Brunelli}\thanks{\texttt{jcbrunelli@gmail.com}}\\[4pt]
\small Departamento de F\'\i sica, CFM Universidade Federal de Santa Catarina,\\
\small Campus Universit\'{a}rio, Trindade, C.P. 476,\\
\small CEP 88040-900, Florian\'{o}polis, SC, Brazil\\[12pt]
{\sc S. Sakovich}\thanks{\texttt{saks@tut.by}}\\[4pt]
\small Institute of Physics, National Academy of Sciences,\\
\small 220072, Minsk, Belarus
}

\date{}

\maketitle

\begin{abstract}
We show how the Yao--Zeng system of coupled short-pulse equations is related to the original short-pulse equation and obtain the correct zero-curvature representation of the Yao--Zeng system via this relationship.
\end{abstract}

\bigskip \bigskip \bigskip

Recently, the short pulse equation (SPE), which can be written as
\begin{equation}
u_{xt} = u + \frac{1}{6} \left( u^3 \right)_{xx}
\label{spe}
\end{equation}
after a scale transformation of its variables, was derived in \cite{SW} as a novel nonlinear wave equation describing the propagation of ultra-short intense infrared pulses in silica optical fibers. In contrast to the celebrated nonlinear Schr\"{o}dinger equation which describes the propagation of sufficiently broad pulses, or slowly varying wave trains whose spectra are narrowly localized around the carrier frequency, the SPE works fine when the pulse contains only several cycles of its central frequency, that is when the pulse spectrum is broad, as was shown in \cite{CJSW} by numerical simulations. Such ultra-short pulses are very important to future technologies of high-speed fiber-optic communication.

Since its first appearance, the SPE \eqref{spe} became a comprehensively studied equation of soliton theory. In \cite{SS1}, it was shown that the SPE is associated with a spectral problem of the Wadati--Konno--Ichikawa type, and a transformation was found which relates the SPE to the celebrated sine-Gordon equation. In~\cite{SS2}, using this transformation, exact soliton solutions of the SPE were derived from known soliton solutions of the sine-Gordon equation. In~\cite{SS3}, the SPE was studied as one of four distinct Rabelo equations of differential geometry \cite{R}. The recursion operator, Hamiltonian structures and conservation laws \cite{B1,B2}, multisoliton solutions \cite{M1} and periodic solutions \cite{P} of the SPE were found and studied as well. By now, nearly a hundred papers has been published on the SPE. Of their number, several papers were dedicated to multi-component, or vector, generalizations of the original scalar SPE \cite{PKB,S,DMH,YZ,M2,F}. Note that the two-component generalizations of the SPE are interesting to nonlinear optics since they can model the propagation of polarized ultra-short pulses---see, for example, \cite{KKS} and references therein.

In the present short note, we consider one of two-component generalizations of the SPE, namely, the Yao--Zeng system (YZS) \cite{YZ}
\begin{equation}
u_{xt} = u + \frac{1}{6} \left( u^3 \right)_{xx} , \qquad v_{xt} = v + \frac{1}{2} \left( u^2 v_x \right)_x .
\label{yzs}
\end{equation}
Our aim is to show how simply this system is related to the original SPE \eqref{spe}, and to obtain via this relationship the correct zero-curvature representation (ZCR) of the YZS \eqref{yzs}. The reason to do so is the following. In \cite{YZ}, a definite first-order linear spectral problem was considered, a hierarchy corresponding to that spectral problem was constructed, and the system \eqref{yzs} was claimed to belong to that hierarchy. However, one can check easily that there must be some errors in matrices of the ZCR of \eqref{yzs} given in \cite{YZ} because the ZCR with those matrices actually does not determine the system \eqref{yzs}. Moreover, by direct analysis of the ZCR, one can prove that the system \eqref{yzs} cannot be associated with the spectral problem considered in \cite{YZ}. Since Yao and Zeng has not corrected their results as yet, the question whether the YZS \eqref{yzs} is integrable or non-integrable deserves a separate study.

The SPE \eqref{spe} possesses the linear spectral problem
\begin{equation}
\Phi_x = X \Phi , \qquad \Phi_t = T \Phi ,
\label{lin1}
\end{equation}
or, equivalently, the ZCR
\begin{equation}
D_t X = D_x T - [ X , T ] ,
\label{zcr1}
\end{equation}
where
\begin{equation}
X =
\begin{pmatrix}
\lambda & \lambda u_x \\
\lambda u_x & - \lambda
\end{pmatrix}
, \qquad
T =
\begin{pmatrix}
\frac{\lambda}{2} u^2 + \frac{1}{4 \lambda} & \frac{\lambda}{2} u^2 u_x - \frac{1}{2} u \\[4pt]
\frac{\lambda}{2} u^2 u_x + \frac{1}{2} u & - \frac{\lambda}{2} u^2 - \frac{1}{4 \lambda}
\end{pmatrix}
,
\label{xt}
\end{equation}
$\Phi$ is a two-component column, $D_x$ and $D_t$ stand for the total derivatives, square brackets denote the matrix commutator, and $\lambda$ is the spectral parameter. If we make the perturbation
\begin{equation}
u \mapsto u + \epsilon z
\label{per1}
\end{equation}
of the dependent variable $u$ in the SPE \eqref{spe} and retain only the terms with $\epsilon^0$ and $\epsilon^1$, we obtain the Pietrzyk--Kanatt\v{s}ikov--Bandelow system (PKBS) \cite{PKB}
\begin{equation}
u_{xt} = u + \frac{1}{6} \left( u^3 \right)_{xx} , \qquad z_{xt} = z + \frac{1}{2} \left( u^2 z \right)_{xx} ,
\label{pkbs}
\end{equation}
which is essentially the perturbed SPE, as was pointed out by Pietrzyk et al. Applying the perturbation \eqref{per1} together with
\begin{equation}
\Phi \mapsto \Phi + \epsilon \Xi , \qquad X \mapsto X + \epsilon A , \qquad T \mapsto T + \epsilon B
\label{per2}
\end{equation}
to the relations \eqref{lin1}, \eqref{zcr1} and \eqref{xt}, and retaining the terms with $\epsilon^0$ and $\epsilon^1$ only, we obtain
\begin{equation}
\begin{gathered}
\Phi_x = X \Phi , \qquad \Phi_t = T \Phi , \\
\Xi_x = A \Phi + X \Xi , \qquad \Xi_t = B \Phi + T \Xi
\end{gathered}
\label{lin2}
\end{equation}
and
\begin{equation}
\begin{gathered}
D_t X = D_x T - [ X , T ] , \\
D_t A = D_x B - [ A , T ] - [ X , B ]
\end{gathered}
\label{zcr2}
\end{equation}
with
\begin{equation}
\begin{gathered}
A =
\begin{pmatrix}
0 & \lambda z_x \\
\lambda z_x & 0
\end{pmatrix}
, \\
B =
\begin{pmatrix}
\lambda u z & \lambda u z u_x + \frac{\lambda}{2} u^2 z_x - \frac{1}{2} z \\[2pt]
\lambda u z u_x + \frac{\lambda}{2} u^2 z_x + \frac{1}{2} z & - \lambda u z
\end{pmatrix}
.
\end{gathered}
\label{ab}
\end{equation}
We can rewrite these relations \eqref{lin2} and \eqref{zcr2} in the block-matrix form as, respectively, the linear spectral problem
\begin{equation}
\Psi_x = M \Psi , \qquad \Psi_t = N \Psi
\label{lin3}
\end{equation}
and the ZCR
\begin{equation}
D_t M = D_x N - [ M , N ]
\label{zcr3}
\end{equation}
of the PKBS \eqref{pkbs}, where
\begin{equation}
\Psi =
\begin{pmatrix}
\Phi \\
\Xi
\end{pmatrix}
, \qquad
M =
\begin{pmatrix}
X & 0 \\
A & X
\end{pmatrix}
, \qquad
N =
\begin{pmatrix}
T & 0 \\
B & T
\end{pmatrix}
,
\label{block1}
\end{equation}
and the matrices $X$, $T$, $A$ and $B$ are given by \eqref{xt} and \eqref{ab}. Note that Pietrzyk et al. constructed their system and its ZCR directly, not using the perturbation \eqref{per1} and \eqref{per2}, and that their ZCR was given in a different form, which is, however, equivalent via a gauge transformation to the one obtained here.

It is easy to see that the YZS \eqref{yzs} and the PKBS \eqref{pkbs} are related to each other by the potential transformation
\begin{equation}
z = c v_x ,
\label{pt}
\end{equation}
where $c$ is any nonzero constant. The choice of $c \neq 1$ can be convenient, as we will see soon. Direct substitution of $z$ \eqref{pt} to $M$ and $N$ \eqref{block1} leads to a ZCR of the YZS, in the sense that the relation \eqref{zcr3} is satisfied by any solution of the YZS. This ZCR, however, is inconvenient because it determines not the YZS \eqref{yzs} itself but a differential consequence of the YZS, namely, the system consisting of the SPE \eqref{spe} and the third-order equation $v_{xxt} = v_x + \frac{1}{2} \left( u^2 v_x \right)_{xx}$. To improve the situation, we obtain an equivalent ZCR of the PKBS, using a gauge transformation of $\Psi$, $M$ and $N$ in \eqref{lin3} and \eqref{zcr3}:
\begin{equation}
\begin{gathered}
\Psi' = G \Psi , \\
M' = G M G^{-1} + ( D_x G ) G^{-1} , \\
N' = G N G^{-1} + ( D_t G ) G^{-1} .
\end{gathered}
\label{gt}
\end{equation}
We take the matrix $G$ in the block form
\begin{equation}
G =
\begin{pmatrix}
I & 0 \\
H & I
\end{pmatrix}
,
\label{g}
\end{equation}
where $I$ is the $2 \times 2$ identity matrix and
\begin{equation}
H =
\begin{pmatrix}
0 & - \lambda z \\
- \lambda z & 0
\end{pmatrix}
.
\label{h}
\end{equation}
Note that the gauge transformation \eqref{gt} with the choice of \eqref{g} and \eqref{h} does not change the blocks $X$ and $T$, but eliminates $z_x$ from $M$ and brings $z_t$ into $N$. As the result, we obtain
\begin{equation}
M' =
\begin{pmatrix}
X & 0 \\
P & X
\end{pmatrix}
, \qquad
N' =
\begin{pmatrix}
T & 0 \\
Q & T
\end{pmatrix}
,
\label{block2}
\end{equation}
where
\begin{equation}
\begin{gathered}
P =
\begin{pmatrix}
0 & 2 \lambda^2 z \\
- 2 \lambda^2 z & 0
\end{pmatrix}
, \\
Q =
\begin{pmatrix}
0 & \lambda u z u_x + \frac{\lambda}{2} u^2 z_x + \lambda^2 u^2 z - \lambda z_t \\[2pt]
\lambda u z u_x + \frac{\lambda}{2} u^2 z_x - \lambda^2 u^2 z - \lambda z_t & 0
\end{pmatrix}
.
\end{gathered}
\label{pq}
\end{equation}
Then we substitute $z$ \eqref{pt} to $M'$ and $N'$ given by \eqref{block2} with \eqref{pq}, choose $c = \frac{1}{2 \lambda}$ to make the matrix $M'$ homogeneous in $\lambda$, replace $v_{xt}$ by $v + \frac{1}{2} \left( u^2 v_x \right)_x$ in the matrix $N'$ (that is, we use $N'' = N' + N_0$ instead of $N'$, where $N_0 = 0$ for all solutions of the YZS), and in this way obtain the linear spectral problem \eqref{lin3} and the ZCR \eqref{zcr3} of the YZS \eqref{yzs} with the matrices (we omit primes)
\begin{equation}
M =
\begin{pmatrix}
X & 0 \\
R & X
\end{pmatrix}
, \qquad
N =
\begin{pmatrix}
T & 0 \\
S & T
\end{pmatrix}
,
\label{block3}
\end{equation}
where $X$ and $T$ are given by \eqref{xt} and
\begin{equation}
R =
\begin{pmatrix}
0 & \lambda v_x \\
- \lambda v_x & 0
\end{pmatrix}
, \qquad
S =
\begin{pmatrix}
0 & \frac{\lambda}{2} u^2 v_x - \frac{1}{2} v \\[2pt]
- \frac{\lambda}{2} u^2 v_x - \frac{1}{2} v & 0
\end{pmatrix}
.
\label{rs}
\end{equation}

The most important difference between our results and the results of Yao and Zeng lies in the block $R$ of the matrix $M$: we have found that $R_{21} = - R_{12} = - \lambda v_x$, while the starting point of Yao and Zeng was the assumption that $R_{21} = R_{12} = \lambda v_x$. However, it is possible to prove by direct computation that, if one changes the sign of $R_{21}$ from ``$-$'' to ``$+$'' in the block $R$ \eqref{rs} of the matrix $M$ \eqref{block3}, not fixing initially the matrix $N$, and requires that the relation \eqref{zcr3} determines the system \eqref{yzs}, then the matrix $N$ does not exist at all. Therefore it is not surprising that the matrix $N$ of Yao and Zeng is different from ours, and that their expressions actually do not determine the system \eqref{yzs}, as one can check easily.

To sum up, we have shown that the YZS is related to the perturbed SPE by the potential transformation and obtained the correct ZCR of the YZS via this relationship. Of course, this means that the YZS \eqref{yzs} is integrable.


\begin{thebibliography}{17}

\bibitem{SW} T. Sch\"{a}fer and C.E. Wayne, Propagation of ultra-short optical pulses in cubic nonlinear media, Physica D 196 (2004) 90--105.

\bibitem{CJSW} Y. Chung, C.K.R.T. Jones, T. Sch\"{a}fer, and C.E. Wayne, Ultra-short pulses in linear and nonlinear media, Nonlinearity 18 (2005) 1351--1374.

\bibitem{SS1} A. Sakovich and S. Sakovich, The short pulse equation is integrable, J. Phys. Soc. Japan 74 (2005) 239--241.

\bibitem{SS2} A. Sakovich and S. Sakovich, Solitary wave solutions of the short pulse equation, J. Phys. A: Math. Gen. 39 (2006) L361--L367.

\bibitem{SS3} A. Sakovich and S. Sakovich, On transformations of the Rabelo equations, SIGMA 3 (2007) 086.

\bibitem{R} M.L. Rabelo, On equations which describe pseudospherical surfaces, Stud. Appl. Math. 81 (1989) 221--248.

\bibitem{B1} J.C. Brunelli, The short pulse hierarchy, J. Math. Phys. 46 (2005) 123507.

\bibitem{B2} J.C. Brunelli, The bi-Hamiltonian structure of the short pulse equation, Phys. Lett. A 353 (2006) 475--478.

\bibitem{M1} Y. Matsuno, Multiloop solutions and multibreather solutions of the short pulse model equation, J. Phys. Soc. Japan 76 (2007) 084003.

\bibitem{P} E.J. Parkes, Some periodic and solitary travelling-wave solutions of the short-pulse equation, Chaos, Solitons \& Fractals 38 (2008) 154--159.

\bibitem{PKB} M.E. Pietrzyk, I. Kanatt\v{s}ikov, and U. Bandelow, On the propagation of vector ultra-short pulses, J. Nonlinear Math. Phys. 15 (2008) 162--170.

\bibitem{S} S. Sakovich, On integrability of the vector short pulse equation, J. Phys. Soc. Japan 77 (2008) 123001.

\bibitem{DMH} A. Dimakis and F. M\"{u}ller-Hoissen, Bidifferential calculus approach to AKNS hierarchies and their solutions, SIGMA 6 (2010), 055.

\bibitem{YZ} Y. Yao and Y. Zeng, Coupled short pulse hierarchy and its Hamiltonian structure, J. Phys. Soc. Japan 80 (2011) 064004.

\bibitem{M2} Y. Matsuno, A novel multi-component generalization of the short pulse equation and its multisoliton solutions, J. Math. Phys. 52 (2011) 123702.

\bibitem{F} B.F. Feng, An integrable coupled short pulse equation, J. Phys. A: Math. Theor. 45 (2012) 085202.

\bibitem{KKS} D.V. Kartashov, A.V. Kim, and S.A. Skobelev, Soliton structures of a wave field with an arbitrary number of oscillations in nonresonance media, JETP Letters 78 (2003) 276--280.

\end{thebibliography}
\end{document}